\documentclass[aps,prl,twocolumn,groupedaddress]{revtex4}
\begin{document}


\title{A Bell Theorem with no locality assumption} 


\author{Charles Tresser}
\email[]{charlestresser@yahoo.com}
\affiliation{IBM, P.O.  Box 218, Yorktown Heights, NY 10598, U.S.A.}


\date{\today}

\begin{abstract}
We prove here a version of Bell Theorem that does not assume locality.  As a consequence classical realism, and not locality, is the common source of the violation by nature of all Bell Inequalities.  
\end{abstract}

\pacs{03.65.Ta}

\maketitle

%
%
%
%
\noindent
\textbf{A)}  \textbf{Bell Inequalities and Bell Theorems:}
 
Consider a sequence of spin-$\frac{1}{2}$ particles, each \emph{prepared} to have positive spin along the arbitrary vector $\vec{a}$ (hence \emph{normalized spin} $s= +1$ along  $\vec{a}$); thus we have a  sequence $\{s_i(\vec{a})=+1\}$ for the normalized spins of the successive identically prepared particles.  If one measures the spins of the particles so prepared along a vector $\vec{b}$, one obtains a sequence $s'_i(\vec{b})$  of normalized spins along $\vec{b}$ such that $s'_i(\vec{b})$ is either +1 or -1 (with probability one). Then Quantum Mechanics (QM) teaches us (and experiments confirm) that the average value $\langle  s_i(\vec{a}) |  s'_i(\vec{b})\rangle$ is equal to $\cos (\vec{a},\,\vec{b})$.  This is the spin-$\frac{1}{2}$ version of the classical Malus law for polarization (see, \emph{e.g.,} \cite{Le Bellac}), which may be better known in terms of probabilities of coincidences:  
\[
\pi (s_i(\vec{a})= s'_i(\vec{b}))=\frac{1+ \langle s_i(\vec{a}) |   s'_i(\vec{b})\rangle  }{2}= \cos ^2 (\frac{(\vec{a},\,\vec{b})}{2})\,.
\]

Following  Bohm's  treatment  \cite{Bohm} of the EPR Paradox \cite{EPR} we deal with \emph{electron-positron pairs} and study the \emph{spins} of the particles using \emph{Measurement Tools} (MTs).  \emph{The pairs are prepared in the  \emph{singlet state}  (\cite{Bohm}, p. 400) whose spin part is rotationally invariant and given along any vector by
$$
\Psi(x_1,x_2)=\frac{1}{\sqrt{2}}(| +\rangle _1\otimes| -\rangle_2-| -\rangle_1\otimes|
+\rangle_2)\,,
$$
so that the total spin is 0 (we are neglecting here a possible phase problem that would not affect the strong inequalities that one obtains at the end)}.  

One particle of each pair is observed using the MT $E$, and the other using the MT $P$,  these MTs being operated by people or smart machines.  Fix an \emph{MT vector} $\vec{x}$, \emph{i.e.,} a vector orthogonal to the axis along which the p articles separate.  One next measures with $X$, which stands for $E$ or $P$, the normalized projection of the spin along $\vec{x}$ of one of the 2 elements of the $i^{\rm{th}}$ pairs from a sequence of pairs prepared to be in the singlet state.  The measured value is denoted by $X_i(\vec{x})$, and also denoted by $\mathcal E_i$ if $X=E$ and by $\mathcal P_i$ if $X=P$.
The total spin zero property of the singlet state then reads: 
\[ 
\textrm{For all }  \vec{x}\textrm{ and all }  i,
\quad E_i(\vec{x})+P_i(\vec{x})\equiv \mathcal E_i+ \mathcal P_i=0\,.
\] 

\smallskip
A \emph{Bell Inequality} (BI) is an inequality among some number of objects such as the average $ \langle   X_i(\vec{x}) |   Y_i(\vec{y}) \rangle$ or the probability of coincidence $\pi(X_i(\vec{x})= Y_i(\vec{y}))$.  Here $X$ and $Y$ each stand for one of $E$ and $P$ while $\vec{x}$ and $\vec{y}$ belong to a collection of names of fixed or variable MT vectors whose size depends on the BI version.  

\smallskip
A \emph{Bell Theorem} (BT) is (depending on the authors):

\noindent
[-] Either the contradiction obtained by replacing the objects such as $ \langle X_i(\vec{x}) |   Y_i(\vec{y}) \rangle$ or $\pi(X_i(\vec{x})= Y_i(\vec{y}))$ in a BI by their values provided by augmenting QM using the collection ${\mathcal{A}}$ of augmentations described next. 

\noindent
[-] And/Or the implications of such a contradiction.

\smallskip
\emph{The collection ${\mathcal{A}}$ of augmentations of QM:} The collection of augmentations of QM that can be used to allow one to formulate a Bell type argument we call  ${\mathcal{A}}$  (for short, we will  write \emph{QM}${\mathcal{A}}$ to mean ``QM augmented by an element of ${\mathcal{A}}$", and often use just ${\mathcal{A}}$ to mean ``any element of ${\mathcal{A}}$").  Two of the most extreme and most common samples from the collection ${\mathcal{A}}$ are described next:

\noindent 
- i) At one end of the spectrum of strengths of the augmentations of QM, ${\mathcal{A}}$ contains the strong hypothesis of Bell in \cite{Bell} (see also \cite{Bell2004}) that there are \emph{Hidden Variables} (HVs) whose statistical properties are the same as for standard QM, and which are \emph{predictive} (\emph{i.e.,} the future values of the observables are determined by the present state of the world, itself described using all the needed variables, including the HVs, even if no one can predict nor access all of the values of the HVs).   

\noindent
- ii) Toward the other end  of the spectrum of strengths, ${\mathcal{A}}$ contains what we call \emph{Classical realism} (also called \emph{contrafactual definiteness} \cite{Stapp1971}). According to this weaker hypothesis (used, \emph{e.g.,} in \cite{Stapp1971} and \cite{Eberhard1977} in the BT context), the statistical properties are the same as for standard QM and (without assuming any predictability) an observable has a value that is well defined whenever the measurement could be made, even if the measurement is not performed (for instance because another measurement is performed instead). 

\smallskip
A BT deals with QM${\mathcal{A}}$, furthermore assuming \emph{locality} in versions prior to the new one to be presented here, and one concludes that at least one of the following choices needs to be made: the augmentation gets disqualified to avoid the contradiction of that BT, or one uses the fact that the formerly known BIs themselves disappear if one drops the assumption of locality. 

\medskip
Too broad a definition of locality could force us to take a stand on whether ``QM has some nonlocal features" in a loose sense.  To avoid the need to face such a choice, in this paper by \emph{locality} we mean that  \emph{if the evaluations of the spin projections are spatially separated then the values of the observables near one MT are not influenced by the settings of the other MT.}  More explicitly, by assuming locality the effect of choosing which measurement is made on any particle of a pair could \emph{not} affect what is observed at spatial distance $\Delta x$ and time difference $\Delta t$ on the other particle if the observations are \emph{spatially separated} (\emph{i.e.,} $c^2\Delta t^2<\Delta x^2$ where $c$ stands for the speed of light). \emph{Causality} means that no information bearing signal can propagate faster than light and it is known that causality is not violated by a failure of locality because, as proved in \cite{Jordan1983}, correlations between the results of measurements on separate subsystems of any quantum system cannot be used to send signals.  \emph{The new BT derivation proposed here does not use  the locality assumption and thus challenges the usual claim that ``BT indicates non-locality"}.

\medskip
\noindent
\textbf{B)}  \textbf{A simple derivation of a BT:} 

One of the simplest previous BT derivations has been adapted from \cite{Stapp1971} and reported by Penrose (as told to him by Mermin) in \cite{Penrose1989}, \cite{Penrose2004}.
We present it in two parts of which only the first will be used to present the new BT.  

- I) We assume henceforth that the particles are fast enough for the observations made at $E$ and $P$ on each pair to be spatially separated, with $E$ and $P$ fixed in the reference frame where the observations are synchronous or approximately so.  Following \cite{Penrose1989}  and \cite{Penrose2004}, suppose that there are:

\noindent
- Two MT vectors for $E$, $|\rightarrow\rangle$ defining  $0^\circ$, and $|\uparrow \rangle$ at  $90^\circ$ (we use here kets for vectors in physical space: the context should tell if a given ket represents such a vector, a spin state along that vector, or possibly both entities),

\noindent
- Two MT vectors for  $P$, $|\nearrow \rangle$ at $45^\circ$ and $|\searrow \rangle$ at $- 45^\circ$.

\smallskip
Take the \emph{actual} settings to be respectively $|\rightarrow \rangle$ for $E$ and $|\nearrow \rangle$ for $P$ for a first run of experiments, and call $\mathcal E=\mathcal E_1,\, \mathcal E_2, \dots$ and $\mathcal P=\mathcal P_1,\, \mathcal P_2, \dots$ the runs respectively registered by $E$ and $P$.  Then QM tells us that the probability $\pi(\mathcal E_i = \mathcal P_i)$ that $\mathcal E_i$ and $\mathcal P_i$ coincide is 
$
\frac{1}{2}(1-\cos (45^\circ ))=0.146...\,,
$
hence \emph{about} $0.15$.  The QM prediction comes from (\emph{e.g.,}) \emph{Wave Packet Reduction} according to which the entangled singlet state becomes the product state 
$\Psi'(x_1,x_2)=|+_{\vec{a}} \rangle _1\otimes| -_{\vec{a}}\rangle_2$ or $\Psi'(x_1,x_2)=|-_{\vec{a}} \rangle _1\otimes| +_{\vec{a}}\rangle_2$ after the $E$ measurement along $\vec{a}=|\rightarrow \rangle$ is performed.  Then $P_i(\vec{a})=-E_i(\vec{a})$ and the conclusion follows from Malus law.

- II) In the previous versions one next  assumes \emph{locality} so that $\mathcal P$  does not depend on the $E$ setting.  Call $\mathcal E'$ the run that would have been registered by $E$ if the alternate MT vector $|\uparrow \rangle$ had been chosen (assuming ${\mathcal{A}}$).  The probability $\pi(\mathcal E'_i =\mathcal P_i)$ of agreement between $\mathcal E'$ and $\mathcal P$ would then have been again equal to $ \frac{1}{2}(1-\cos (45^\circ ))=0.146...$.  Next,  if the $E$ settings had been $|\rightarrow \rangle$  as initially decided, but the $P$ settings had been $|\searrow \rangle$, then the run at $E$ would have been $\mathcal E$ as before, using locality again.  Denoting by  $\mathcal P'$ the runs that would have been registered by $P$ with the new setting of its MT vector  (assuming ${\mathcal{A}}$), the expectation of coincidence $\pi(\mathcal E _i=\mathcal P'_i)$ between $\mathcal E_i$ and  $\mathcal P'_i$ would have been again
$
\frac{1}{2}(1-\cos (45^\circ ))=0.146...\,.
$
The punch line is then the difference between two ways of comparing the runs $\mathcal E'$ and  $\mathcal P'$:  

\smallskip
[-] On the one hand we have the  \emph{Bell Inequality}
\[
(*)\quad \pi(\mathcal E'_i = \mathcal P_i)+\pi(\mathcal P_i = \mathcal E_i)+\pi(\mathcal E_i = \mathcal P'_i)\geq \pi(\mathcal E'_i = \mathcal P'_i)\,.
\]

\smallskip
\emph{Proof of $(*)$.} All entries being binary, for $\mathcal E'_i$ and $\mathcal P'_i$ to agree requires at least one of the equalities $\mathcal E'_i=\mathcal P_i$, $\mathcal P_i=\mathcal E_i$, or $\mathcal E_i=\mathcal P_i$ to hold true.
 
\medskip
For now we notice that for the angles that we have chosen, the Bell Inequality $(*)$  yields the upper bound 0.45=0.15+0.15+0.15 for the probability $\pi(\mathcal E'_i = \mathcal P_i')$.

\smallskip
[-] On the other hand computing $\pi(\mathcal E'_i = \mathcal P'_i)$ directly according to QM${\mathcal{A}}$ yields the equality $\frac{1}{2}(1-\cos (135^\circ ))=0.854...$, contradicting the former $0.45$ upper bound: this is the BT's contradiction that we were aiming at. 

\medskip
The conclusion of  \emph{Bell's Theorem} is then:

\noindent
 \emph{One at least of ${\mathcal{A}}$ or locality is false in order to explain why the statistical properties from QM violate the BI.}

\smallskip
It is locality that plays the role of the usual suspect (as clearly stated, \emph{e.g.,} by Penrose in \cite{Penrose2004}, p. 589, and by Aspect in \cite{Bell2004}, pp. xxv-xxvi, following a tradition going back to \cite{Bell}).  It is true that without locality the expression $\pi(\mathcal E'_i = \mathcal P'_i)$ makes no sense and the conclusion vanishes, but this hardly proves that locality is the essential hypothesis, and this paper precisely proves that it is not. 

\bigskip
\noindent
\textbf{C)}  \textbf{A Bell Theorem with no locality assumption:} 

\smallskip
- a)  \emph{A new BI.}  

Only $|\nearrow \rangle$ will ever be used at $P$ while we shall continue to consider both $|\rightarrow\rangle$ and $|\uparrow \rangle$ at $E$ in this section.  Furthermore  we assume that measurements are made of $\mathcal P$ and $\mathcal E$ while $\mathcal E'$ is inferred to \emph{make sense}, \emph{i.e.,} do not depend on choices that have not yet been specified (otherwise speaking to be well defined and have well defined (albeit unknown) values) by using the augmentation of QM by ${\mathcal{A}}$.  One could choose instead to not measure either of $\mathcal E$ and $\mathcal E'$, and use the augmentation of QM by ${\mathcal{A}}$ to infer that both of $\mathcal E$ and $\mathcal E'$ have well defined values: this will be done in the Appendix (see point a5)) in order to focus on a single BT in the main text.  Whenever time ordering is considered in a Lorentz frame between two space-time events, the time that the two signals travel at light speed to an observer in that frame is taken into account. We will use the following:
 
\smallskip
\noindent
\textbf{Effect After Cause Principle (\emph{EACP}):} \emph{For any Lorentz observer and for any $\mathcal X$ in $\{ \mathcal E,\, \mathcal E',\, \mathcal P  \}$, a value $\mathcal X_i$ does not change from any cause that happens after $\mathcal X_i$ has been either measured or inferred  to make sense in view of QM${\mathcal{A}}$ for that observer.}

\smallskip
\noindent
\textbf{EACP vs locality Lemma.} \emph{The EACP is different from locality.}

\noindent
See the Appendix for three proofs of this statement. 

\smallskip
\noindent
\textbf{Definition.}  We call an \emph{$E$-$P$ observer} a Lorentz observer for whom measurements at $E$ occur before measurements at $P$ for each pair, while a \emph{$P$-$E$ observer} travels in the opposite direction.  The $E$-side sequences depend on $\mathcal P_i$ for $P$-$E$ observers and our $E$-$P$ observers will not even consider the $P$ side. 
 
\smallskip
\noindent
\textbf{New Bell Inequality.} \emph{The sum of all three probabilities of coincidences for the chosen triple of pairs must be over $\pi(\mathcal P_i = \mathcal P_i)\equiv 1$, or}
$$
(**)\qquad \pi(\mathcal P_i = \mathcal E_i)+\pi(\mathcal E_i = \mathcal E'_i)+\pi(\mathcal E'_i = \mathcal P_i)\geq 1\,.
$$

\emph{Proof of $(**)$.}  Collecting the results from an $E$-$P$ observer (for $\pi(\mathcal E_i = \mathcal E'_i)$) and a $P$-$E$ observer (for the two other probabilities), it follows readily from the EACP that  the three sequences $\mathcal E$, $\mathcal E'$, and $\mathcal P$ involved in $(**)$ make sense.  All entries being binary, for any $\mathcal P_i$ to agree with itself, one needs at least one of the equalities $\mathcal P_i=\mathcal E_i$, $\mathcal E_i=\mathcal E'_i$, or $\mathcal E'_i=\mathcal P_i$ to hold true.

\bigskip
- b)  \emph{From the new BI to a BT.}  

\smallskip
\noindent
{\bf No Correlation Lemma.} \emph{The sequences $\mathcal E$  and $\mathcal E'$ are not correlated,} \emph{i.e.,} 
\[
(\circ)\quad  \langle\mathcal E_i  |   \mathcal E'_i  \rangle=0 \quad\textrm{or equivalently}\quad\pi(\mathcal E_i = \mathcal E'_i)=\frac{1}{2}\,.
\] 

\noindent
\emph{Proof of the No Correlation Lemma.} Using the EACP and the conclusion deduced from it in the proof of $(**)$ that ``the three sequences $\mathcal E$, $\mathcal E'$, and $\mathcal P$ involved in $(**)$ \emph{make sense}",  we notice that only the orientation of the angle  $(|\rightarrow\rangle, |\uparrow \rangle)$ at $E$ could matter for an $E$-$P$ observer, so that $(\circ)$ follows from invariance under parity without assuming locality. 

To  see the role of parity, we introduce the further vector $-|\uparrow \rangle$ to which would correspond the sequence $\mathcal E''$ such that  $\mathcal E''_i\equiv 1-\mathcal E'_i$.  Since $\pi(\mathcal E_i = \mathcal E'_i)+\pi(\mathcal E_i = \mathcal E''_i)=1$ it only remains to prove these two probabilities to be equal to each other.  We use here sequences that are possibly unknown, but known to be well defined:

- one, $\mathcal E$, is known by direct measurement, 

- the other $\mathcal E'$,  can be inferred to be unknown but well defined by an $E$-$P$ observer using QM${\mathcal{A}}$. 

\noindent
The only thing that could generate inequalities is the difference in the orientations of the angles $(|\rightarrow \rangle, |\uparrow \rangle )$ and $(|\rightarrow \rangle, -|\uparrow \rangle)$. Equality thus follows from parity invariance.

\smallskip
\noindent
\textbf{New Bell Theorem:} \emph{We can use the triplet}

\noindent
$( \pi(\mathcal P_i = \mathcal E_i),\pi(\mathcal E'_i = \mathcal P_i),  \pi(\mathcal E_i = \mathcal E'_i))=(0.15, 0.15, 0.5)\,$ 

\noindent
\emph{of probabilities of coincidences in $(**)$ specialized to the chosen angles,  yielding $0.8\geq 1$ as the false inequality that is the contradiction that we seek.  Since no locality  assumption  is used in the derivations  of the triplet and $(**)$, only the \emph{counter-natural} character of the \emph{gedanken} experiment (\emph{i.e.,} its dependence upon observable values that cannot be measured and are thus inferred to make sense assuming ${\mathcal{A}}$) can be the cause of the contradiction between Quantum Mechanics statistics and the new Bell Inequality}.  

\smallskip
\emph{Proof of the New Bell Theorem.}  Again we assume the EACP (in full compatibility with all experiments made so far) and use $(**)$ and the proof of it that is provided above.
 
 \smallskip
- 1) After measurements are made using $P$, a $P$-$E$ observer obtains  that:

\noindent
- e1) $\pi (\mathcal P_i=\mathcal E_i)$ is about $0.15$ by QM, or by direct observation after measurements are also made using $E$,

\noindent
 - e2) $\pi (\mathcal P_i=\mathcal E'_i)$ is about $0.15$ by QM${\mathcal{A}}$.
 
 The deductions made in e1) and e2) using QM${\mathcal{A}}$ go as follows:  By Wave Packet Reduction (for instance), the spin state of second particle (the particle on the $E$ side) becomes 
\[
\Psi(x_2)=|\mathcal P_i \rangle _1\otimes| -\mathcal P_i \rangle_2 
\]
along the vector $|\nearrow \rangle$ (along which the sequence $\mathcal P$ is measured) as soon as the measurement of $\mathcal P_i$ is made on the $P$ side.  Hence the second particle gets into a spin state prepared to be $ |-\mathcal P_i \rangle $ along the vector $|\nearrow \rangle$ (as revealed by using the information obtained on the $P$ side)  so that both of the two probabilities $\pi (\mathcal P_i=\mathcal E_i)$ and $\pi (\mathcal P_i=\mathcal E'_i)$ are equal to about $0.15$ by a simple application of the spin-$\frac{1}{2}$ version of Malus law for polarization as we have recalled it, under the assumption found in both of the rather extreme elements  of ${\mathcal{A}}$ described above that the statistics of QM${\mathcal{A}}$ is the same as the statistics of standard QM.

 \smallskip
- 2) An $E$-$P$ observer infers: 

\noindent
 - e3) $\pi (\mathcal E_i=\mathcal E'_i)= 0.5$ on the $E$ side by the No Correlation Lemma. 

\smallskip
Assembling the conclusions  e1), e2), and e3) from the two (strongly) asynchronous frames (\emph{e.g.,} in the Lorentz frame of the experiment since the outcomes cannot change according to the Lorentz frame) one obtains  the expected triplet evaluation for the three probabilities: $(0.15, 0.15, 0.5)$.  Together with $(**)$, this evaluation provides us with the impossible inequality $0.8\geq 1$. The rest of the theorem follows from checking the hypotheses that are used in the proof that we have given of the contradiction.  This concludes the proof of the New Bell Theorem.

\medskip
Our New Bell  Theorem admits the following immediate corollary that we will use as our conclusion:

\medskip
\noindent
\textbf{Corollary:} \emph{It is the counter-natural character of the reasoning, as permitted by any form of ${\mathcal{A}}$, and not locality, that is the \textbf{only} problem common to \textbf{all} the versions of Bell Theorem, since one needs to assume some form of ${\mathcal{A}}$ in any argument of Bell's type.   Thus Non-locality is not needed (in some circles, one would say that it can be disposed of using Occam's razor).}

\appendix
\medskip
\noindent
\textbf{D)}  \textbf{Appendix:} 

- a1) For an analysis of the essential equivalence of all the BIs that use some locality asumptions, see  \cite{Fine1982a} and \cite{Fine1982b}.  The probabilistic approach of Fine was later continued by Pitowsky (see \cite{Pitowsky1994}, \cite{Pitowsky2001} and references therein) who identified BI's as examples of inequalities going back to Boole \cite{Boole1854}.  This lets us understand BIs as resulting from a classicality assumption.

\smallskip
- a2) Assuming locality, a simple proof of $(\circ)$ (that readily follows from the arguments of Bell in \cite {Bell}) was given by Stapp (see equation (7) in \cite{Stapp1971}).

\smallskip 
- a3) The inequality $(**)$ is a special case of  (2a) in \cite{Fine1982b}, established there assuming locality (see also (4) in \cite{Pitowsky1994}).  

\smallskip
- a4) We could not have used $(*)$ instead of $(**)$, but we could have used the following weaker but new form of BI:
\[
\min [ \max (\pi(\mathcal P_i = \mathcal E_i),\, \pi(\mathcal P_i = \mathcal E'_i))]\geq \frac{1- \pi(\mathcal E'_i = \mathcal E_i)}{2}\,.
\]

\smallskip
- a5) If no measurement is made on the $E$ side, one can still prove $(\circ)$.  We start as in the case when one measures $\mathcal E$.  Like in that case (\emph{cf.} the main text) the only thing that could generate inequalities is the difference in the orientations of the angles $(|\rightarrow \rangle, |\uparrow \rangle )$ and $(|\rightarrow \rangle, -|\uparrow \rangle)$.  But now, $\mathcal E$ and $\mathcal E'$ have symmetrical roles because none of the sequences gets to be measured.  Both of these sequences are now inferred to have definite value by  an $E$-$P$ observer on the basis of QM${\mathcal{A}}$.  Reversing the order of the elements of the second pair one would find $|\rightarrow \rangle$ in the role payed by $|\uparrow \rangle$ in the first pair, from which the equality follows immediately for an $E$-$P$ observer assuming isotropy (no need of parity invariance in this case): this provides us with e3).  The proof of e2) is unchanged and e1) is proved like e2).  We thus recover a BT when no measurement is made on the $E$ side. 

\smallskip
- a6) \emph{Three proofs of EACP$\not =$locality.} Let us not assume locality.  Then in the setting for a BI presented in part \textbf{B)}, the EACP would still allow a $P$-$E$ observer to make sense of $\pi ( \mathcal P_i= \mathcal E_i)$ and of  $\pi ( \mathcal P_i= \mathcal E'_i)$ as was done in the proof of the New Bell Theorem to justify e1) and e2) and, in the same way, an $E$-$P$ observer to make sense of $\pi ( \mathcal E_i= \mathcal P_i)$ and of  $\pi ( \mathcal E_i= \mathcal P'_i)$.  However one cannot extract $\pi ( \mathcal E'_i= \mathcal P'_i)$ from augmenting QM by ${\mathcal{A}}$ without making a more precise assumption, to the contrary of what happens if on assumes locality.   This concludes a first proof that  EACP$\not =$locality.  To get another proof,  notice that if one assumes non-locality in its strongest form together with EACP, one gets that $\pi ( \mathcal E'_i= \mathcal P'_i)=0.5$ (meaning statistical independence) instead of the $0.85$ that was obtained in section \textbf{B)} under the usual locality assumption, enough though, since $0.5>0.45$, to get a BT.  To get a third proof,  notice that by violating the EACP  when measurements are made, one would (easily) enable super-luminal signaling, to the contrary of what happens with any locality violation in view of \cite{Jordan1983}.  Details are left to the reader. 

\smallskip
\noindent
\textbf{Remark:} \emph{If one uses a broader and more common definition of locality than the one that we have chosen, things get even simpler since, as Richard Frieberg puts it ``the difference between EACP and locality is that locality in a relativistic context means \emph{(in its usual broader sense)} that no influence at all can be exerted across spacelike separation".}


\begin{acknowledgments}
\textbf{Acknowledgments:} 
I  thank M. le Bellac, O. Cohen, P. Coullet, R. Griffiths, G. t'Hooft, L. Horwitz, D. Mermin, D. Ostrowsky,  I. Pitowsky, Y. Pomeau,  and  L. Vaidman  for questions or suggestions after reading or listening to versions of this work., and I cannot find the words to thank Arthur Fine, Richard Friedberg, Pierre Hohenberg and Edward Spiegel for their patience, critics, advices, and encouragements. 
\end{acknowledgments}

%
%
%
%

\end{document}